\documentclass[final,1p,times]{elsarticle} 

\usepackage{graphicx}
\usepackage{amssymb} 
\usepackage{amsmath} 
\usepackage{amsthm} 
\usepackage{lineno}

\journal{Nuclear Physics A} 

\begin{document}

\begin{frontmatter} 

  \title{Effect of longitudinal fluctuation in event-by-event (3+1)D hydrodynamics}

  \author[ustc,lbnl]{Longgang Pang}
  \author[ustc]{Qun Wang}
  \author[ccnu,lbnl]{Xin-Nian Wang}

  \address[ustc]{Interdisciplinary Center for Theoretical Study and Department of Modern Physics, University of Science and Technology of China, Hefei 230026, China}
 \address[ccnu]{Key Laboratory of Quarks and Lepton Physics (MOE) and Institute of Particle Physics, Central China Normal University, Wuhan 430079, China}
  \address[lbnl]{Nuclear Science Division, MS 70R0319, Lawrence Berkeley National Laboratory, Berkeley, CA 94720}

  \begin{abstract} 
    Hadron spectra and elliptic flow in relativistic heavy-ion collisions are studied in event-by-event (3+1)D ideal hydrodynamic simulations with fluctuating initial conditions given by the AMPT Monte Carlo model. Both the coherent soft gluon production from wounded nucleons and incoherent mini-jet production in semi-hard parton scatterings are considered. These partons take part in parton cascade and are assumed to be locally thermalized. They provide the fluctuating initial conditions for hydrodynamic simulations with a Gaussian smearing. Effects of both transverse and longitudinal fluctuations are studied.
The initial fluctuations along rapidity direction lead to expanding hot spots in longitudinal direction, which will reduce elliptic flow and the yield of particles at high transverse momentum.  The intrinsic correlation introduced in the fluctuating initial conditions is also found to influence the di-hadron correlation of the final hadrons.
  \end{abstract} 

\end{frontmatter} 


\section{Fluctuating initial conditions from AMPT}
Fluctuation in the initial condition for hydrodynamic study of heavy-ion collisions have been found necessary for the description of
the anisotropic flow of final hadrons. We have carried out event-by-event (3+1)D ideal hydrodynamic simulations \cite{Pang:2012he} 
with the local initial energy-momentum tensor in each hydrodynamic cell given by AMPT \cite{Zhang:1999bd} model.
AMPT uses HIJING Monte Carlo Model \cite{Wang:1991hta, Gyulassy:1994ew} for the initial parton production from hard
scatterings and excited strings from soft interactions.
The number of excited strings in each event is equal to participant nucleons.
The number of mini-jets per binary nucleon-nucleon collisions follows a Poisson distribution with 
the average value given by mini-jet cross section, which depends on the colliding energy and the impact parameter 
through an impact-parameter dependent parton shadowing \cite{Wang:1991hta} in a nucleus.
HIJING uses the Glauber model to determine the number of participants and binary nucleon-nucleon 
collisions with the Wood-Saxon nuclear distribution. Both soft partons from string melting and mini-jet partons will take part 
in parton cascade for a short period. In this model, the total local energy-momentum 
density of partons and its fluctuations will be determined by the number of participants, binary
nucleon-nucleon collisions, number of mini-jets per nucleon-nucleon collision, the fragmentation
of excited strings, and the parton cascading. 

The $4$-momenta and space coordinates of partons from the AMPT model according to the above description will be used 
to calculate the local energy-momentum tensor for the initial conditions of our
event-by-event (3+1)D hydrodynamic simulations. Its value in each grid cell is approximated by a Gaussian 
distribution in invariant-time coordinates,
\begin{equation}
  T^{\mu\nu} (\tau_{0},x,y,\eta_{s}) = K\sum_{i}
  \frac{p^{\mu}_{i}p^{\nu}_{i}}{p^{\tau}_{i}}\frac{1}{\tau_{0}\sqrt{2\pi\sigma_{\eta_{s}}^{2}}}\frac{1}{2\pi\sigma_{r}^{2}}
  \times \exp \left[-\frac{(x-x_{i})^{2}+(y-y_{i})^{2}}{2\sigma_{r}^{2}} - \frac{(\eta_{s}-\eta_{i s})^{2}}{2\sigma_{\eta_{s}}^{2}}\right],
  \label{eq:Pmu}
\end{equation}
where $p^{\tau}_{i}=m_{iT}\cosh(Y_{i}-\eta_{i s})$, $p^{x}_{i}=p_{i x}$, $p^{y}_{i}=p_{i y}$ 
and $p^{\eta}_{i}=m_{i T}\sinh(Y_{i}-\eta_{i s})/\tau_{0}$ for parton $i$, which runs over all partons produced in the AMPT 
model simulations. 
We have chosen $\sigma_{r}=0.6$ fm, $\sigma_{\eta_{s}}=0.6$ in our calculations.
The transverse mass $m_{T}$, 
rapidity $Y$ and  spatial rapidity $\eta_{s}$ are calculated from the parton's $4$-momenta and spatial coordinates. 
Note that the Bjorken scaling assumption $Y=\eta_{s}$ is not used here because of early parton cascade before the initial time and the uncertainty
principle applied to the initial formation time in AMPT. 
We used $K=1.45$ and $\tau_0=0.4$ fm in Au+Au collisions at $\sqrt{s}_{NN}=200$ GeV in the following calculations 
which gives good description of the multiplicity density of produced hadrons at central rapidity.

\section{The effects of longitudinal fluctuations on $p_{T}$  spectra and elliptic flow}

To investigate the effect of longitudinal fluctuations on hadron spectra and elliptic flow, we compare the (3+1)D hydrodynamic calculations 
using the full AMPT initial conditions with that using a tube-like smooth initial longitudinal distribution.
In the event-by-event tube-like initial condition, we take the energy density and transverse flow velocity from AMPT results in the central rapidity region,
and assume the fluctuation to be the same along longitudinal direction with an envelope distribution function,
\begin{equation}
  H(\eta)= \exp\left[-\theta(|\eta|-\eta_{0})(|\eta|-\eta_{0})^2/2\sigma_{w}^{2}\right],
  \label{eq:tube}
\end{equation}
in rapidity. The length of the plateau $\eta_0$ in the center rapidity region and the width of the Gaussian fall-off $\sigma_w$ at large rapidity 
are adjusted to fit the multiplicity distribution of charged hadrons.  
In another comparison, we also calculate the hadron spectra and elliptic flow for one shot tube initial condition,
which is averaged over many AMPT initial conditions but rotated by an angle in each event to a common participant plane.
We refer to this as one-shot-tube initial condition since in longitudinal direction the energy density and flow velocity also have a smooth 
tube-like distribution as defined in Eq. \ref{eq:tube}.

\begin{figure}[ht]
  \begin{center}
    \includegraphics[width=0.8\textwidth]{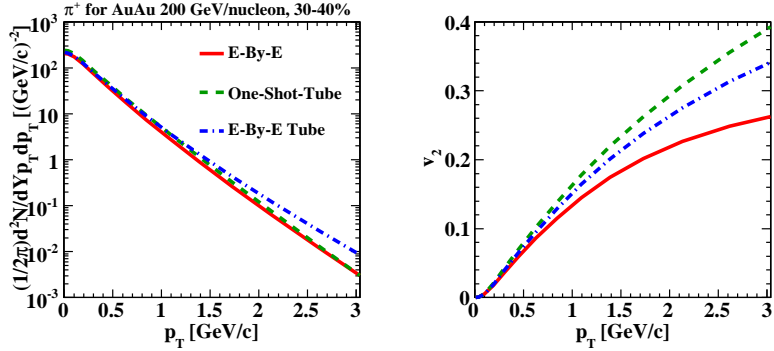}
  \end{center}
  \caption{(Color online) The $p_{T}$  spectra (left panel) and elliptic flow (right panel) for charged pion given by hydrodynamic simulations at $30-40\%$ semi-central Au+Au collisions for full AMPT initial conditions (solid lines), event-by-event tube initial conditions (dot-dashed lines) and one-shot-tube initial conditions (dashed lines). }
  \label{fig:fluctuations}
\end{figure}

Shown in Fig. \ref{fig:fluctuations} are the transverse momentum spectra (left panel) and elliptic flow (right panel) for $\pi^+$ in $30-40\%$ semi-central Au+Au collisions at the RHIC energy $\sqrt{s}_{NN} = 200 $ GeV with the full AMPT initial conditions (solid lines), event-by-event tube initial condition (dot-dashed lines) and one-shot-tube initial conditions (dashed).
The event-by-event fluctuations in the tube-like AMPT initial conditions significantly reduce the elliptic flow of final hadrons respect to the event plane as compared to the one-shot-tube initial conditions.
The yield of high $p_{T}$  particles from event-by-event tube initial conditions on the other hand is increased due to the fluctuations (both in energy density and flow velocity) in the transverse direction as compared to the spectra from one-shot-tube initial conditions.
Similar results were found in (2+1)D \cite{Qiu:2011iv} and (3+1)D \cite{Schenke:2010rr} hydrodynamic simulations.
This can be understood as the fast isotropic expansion of each hot spot in the transverse direction at the the early time of the hadrodynamic evolution.

Fluctuations in the longitudinal direction in the full AMPT initial conditions also lead to hot spots in the longitudinal direction.
Expansion of such longitudinal hot spots will dissipate more transverse energy in the longitudinal direction.
This in turn decreases the elliptic flow at large $p_{T}$  as compared to the event-by-event tube initial condition.
The yield of high $p_{T}$  particles are reduced due to the expansion of longitudinal hot spots from full AMPT initial conditions as compared to event-by-event tube-like AMPT initial conditions. 

Since the anisotropic flow is used to extract the transport coefficients (such as shear viscosity) from comparisons between experimental data and viscous hydrodynamics, the inclusion of longitudinal fluctuations in the hydrodynamic calculation will be necessary for more quantitative studies.

\section{The effects of longitudinal fluctuations on di-hadron correlation}

To study the effect of longitudinal fluctuations on the di-hadron correlation, we compare the di-hadron correlation for charged hadrons in (3+1)D hydrodynamic simulations with the full AMPT initial conditions with the event-by-event tube-like AMPT initial conditions.
The two dimensional di-hadron correlation is defined as,
\begin{equation}
  C12(\Delta \eta, \Delta \phi) = \frac{<N_{t}(\eta_1, \phi_1) N_{a}(\eta_2, \phi_2) >_{same}}{<N_t(\eta_1, \phi_1) N_a(\eta_2, \phi_2)>_{mixed}} 
  \label{eq:c12}
\end{equation}
where $N_t $ and $N_a $ are the number of trigger and associate particles in two bins centered at $(\eta_1, \phi_1)$ and $(\eta_2, \phi_2)$. Particles with the same rapidity $\Delta\eta = |\eta_1-\eta_2|$ and azimuthal separation $ \Delta\phi=|\phi_1-\phi_2|$ are averaged to increase statistics.
The range of $\Delta \eta$ and $\Delta \phi$ are extrapolated in the plot by symmetries of the correlation function.
Correlation of particles from the same event is divided by that from mixed events to take  into account of the finite rapidity window.
In calculating the correlation for mixed events in the denominator, we rotate each event by a random angle in azimuthal direction 
to remove the background correlation coming from the same reaction plane used in AMPT model.

\begin{figure}[ht]
  \begin{center}
    \includegraphics[width=0.8\textwidth]{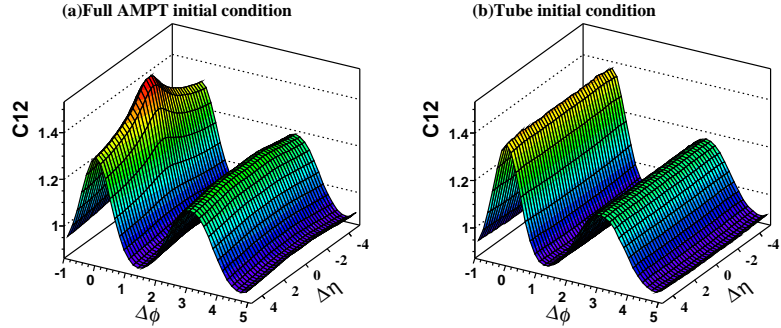}
  \end{center}
  \caption{(Color online) Di-hadron correlations for charged hadrons calculated from (3+1) hydrodynamics for $30-40\%$ semi-central Au+Au collisions at RHIC energy $\sqrt{s}_{NN}=200$ GeV, from (a) full AMPT initial conditions and (b) event-by-event tube-like AMPT initial conditions. The transverse momentum cut-off for both trigger and associate particles are $2<p_{T}< 3$ GeV/c. }
  \label{fig:c12}
\end{figure}
We show in Fig. \ref{fig:c12} the di-hadron correlation for charged hadrons in (3+1)D hydrodynamic simulation from full AMPT initial conditions and event-by-event tube-like initial conditions. 
The di-hadron correlation is flat in rapidity direction for tube-like initial conditions as compared to the full AMPT initial condition.
This proves that the hydrodynamic simulation is not able to produce a near-side peak for low $p_{T}$ particles
if no fluctuations and intrinsic correlations in longitudinal direction are introduced in the beginning of the hydrodynamic evolution. 

\section{Summary}

Using fluctuating initial conditions from the AMPT Monte Carlo model for event-by-event (3+1)D ideal hydrodynamic simulations,
we introduce both fluctuations and intrinsic correlations in rapidity direction in the initial condition.
Effects of longitudinal fluctuations (both in energy density and flow velocity) on hadron spectra, elliptic flow and di-hadron correlation are studied in hydrodynamic simulations from full AMPT initial condition, event-by-event tube-like initial condition and one-shot-tube initial condition.
We showed that fluctuations in rapidity direction reduce noticeably the elliptic flow at large transverse momentum, which brings more
 constraint on extracting $\eta/s$ from comparison between the experimental data with viscous hydrodynamic calculations.
The initial longitudinal fluctuations with intrinsic correlation built in is found to be important to understand the near-side peak of 
the di-hadron correlations for final particles at low $p_{T}$.


\section*{References}

\begin{thebibliography}{00} 
    \bibitem{Pang:2012he} 
      L.~Pang, Q.~Wang and X.~-N.~Wang,
      Phys.\ Rev.\ C {\bf 86}, 024911 (2012)
      [arXiv:1205.5019 [nucl-th]].

      \bibitem{Zhang:1999bd}
	B.~Zhang, C.~M.~Ko, B.~-A.~Li, Z.~-w.~Lin,
	Phys.\ Rev.\  {\bf C61}, 067901 (2000).
	[nucl-th/9907017].

	\bibitem{Wang:1991hta}
	  X.~-N.~Wang, M.~Gyulassy,
	  Phys.\ Rev.\  {\bf D44}, 3501-3516 (1991).

	  \bibitem{Gyulassy:1994ew}
	    M.~Gyulassy and X.~-N.~Wang,
	    Comput.\ Phys.\ Commun.\  {\bf 83}, 307 (1994)
	    [nucl-th/9502021].

	    \bibitem{Qiu:2011iv} 
	      Z.~Qiu and U.~W.~Heinz,
	      Phys.\ Rev.\ C {\bf 84}, 024911 (2011)
	      [arXiv:1104.0650 [nucl-th]].


	      \bibitem{Schenke:2010rr} 
		B.~Schenke, S.~Jeon and C.~Gale,
		Phys.\ Rev.\ Lett.\  {\bf 106}, 042301 (2011)
		[arXiv:1009.3244 [hep-ph]].


		\end{thebibliography}
		\end{document}